\documentclass[prd,showkeys,floatfix,nofootinbib,
               preprint,12pt,
               tightenlines,fleqn]{revtex4} %for preprint

\usepackage{amsmath,amssymb,revsymb,graphicx,dcolumn}
\usepackage{leftidx}
\newcommand{\beq}{\begin{equation}}
\newcommand{\eeq}{\end{equation}}
\newcommand{\beqa}{\begin{eqnarray}}
\newcommand{\eeqa}{\end{eqnarray}}
\newcommand{\bsubeqs}{\begin{subequations}}
\newcommand{\esubeqs}{\end{subequations}}

\begin{document}
\title[]
      {Local thermal observables in spatially open FRW spaces\vspace*{5mm}}
\author{Slava Emelyanov}
\email{viacheslav.emelyanov@physik.uni-muenchen.de}
\affiliation{Arnold Sommerfeld Center for
Theoretical Physics,\\
Ludwig Maximilian University (LMU),\\
80333 Munich, Germany\\}

\begin{abstract}
\vspace*{2.5mm}\noindent
Certain local thermal observables are considered in well-known examples of spatially open
FRW spaces: Milne, open de Sitter and anti-de Sitter as well as Einstein static universes. 
Another value for fixing the ambiguity in defining the Wick square and, hence, the local
temperature is motivated in the last example. Physical consequences of that choice are discussed
for static and conformal vacua in those spaces.
\end{abstract}

%\pacs{98.80.Es, 98.80.Cq, 04.20.Cv}


\maketitle

\section{Introduction}

In this paper I will explore, in particular, thermal characteristics of states considered in~\cite{Emelyanov}.
A framework employed below is based on the idea that one can construct microscopic quantities from 
the field products and its derivatives which are sensitive to thermal properties of a certain quantum 
state~\cite{Buchholz&Ojima&Roos,Buchholz}. Among of these local thermal observables are a local 
temperature and a thermal energy-momentum tensor. Thus, computing them in a quantum state under
consideration and in a some reference thermal one, one may decide to which extent it is legitimate to 
ascribe macroscopic thermal observables to it as well as whether it corresponds to a local thermal 
equilibrium. 

Although there exists in general no global thermal states in curved spacetimes to be taken as reference 
ones, it is not the case in the problem under scrutiny. Specifically, I shall consider observers moving along 
geodesics corresponding to the integral curves of the conformal Killing vector field in spatially open FRW 
spaces among of which are Milne, open de Sitter (dS), anti-de Sitter (AdS) spaces as well as Einstein 
static universe (ESU).\footnote{Spacetime is open or closed if the spatial section is $\mathbf{H}^3$ or 
$\mathbf{S}^3$, respectively.} Therefore, a conformal Kubo-Martin-Schwinger (KMS) state~\cite{Haag} is 
chosen in all of these cases (except the last one, wherein it is just a KMS state) as the reference thermal one.

In Section \ref{sec:milne}, I consider the contracting and expanding Milne universes. An alternative 
quantization being equivalent to the standard one and relation with the conformal vacuum are studied. 
Thermal properties of known states are discussed in the framework briefly outlined above. In Section 
\ref{sec:ads}, I deal with AdS spacetime, wherein an alternative quantization of the conformal non-interacting 
scalar field on the AdS hyperboloid will be demonstrated that is unitary equivalent to the standard 
one~\cite{Avis&Isham&Storey} and the existence of which has been motivated in~\cite{Emelyanov}. 
Along the line of Section \ref{sec:milne}, a discussion of the thermal properties of the static and conformal 
vacua is presented. In Section \ref{sec:discussion}, I discuss thermal local observables in the cases of 
open dS space as well as open Einstein static universe and compare them with those for closed dS space 
and the closed ESU. In Section \ref{sec:conclusions}, I provide final concluding remarks.

The sign convention for the metric tensor as well as the definition of the Riemann tensor are
the same as in~\cite{Emelyanov}. The fundamental constants are set to unity throughout this paper.

\section{Milne spacetime}
\label{sec:milne}

The Milne universe is a subspace of Minkowski spacetime lying in the future lightcone originating from a given 
point O of the manifold. Its line element has the form of the open FRW universe with the scale factor exponentially 
growing with the conformal time, i.e.
\beqa
ds^2 &=& a^2(\bar\eta)\big(d\bar\eta^2 - d\bar\chi^2 - \sinh^2\bar\chi d\bar\Omega^2\big)\,,
\eeqa
where $a(\bar\eta) = e^{\bar\eta}$ is the scale factor and $d\bar\Omega^2 = d\bar\theta^2 + \sin^2\bar\theta d\bar\varphi^2$
element of solid angle. By the same metric with a reversed direction of $\bar\eta$ one can cover the past lightcone 
of the origin O, wherein now $a(\bar\eta)$ approaches zero for $\bar\eta \rightarrow + \infty$. The future and past 
lightcones will be called the upper and lower Milne universes in the following, respectively.

\subsection{Minkowski modes in Milne spacetime}

To simplify calculations of the Wick square (see below) in the conformal KMS state, one has to expand the quantum 
field through modes which when rescaled becomes positive frequency ones with respect to the conformal Killing 
vector $\partial_{\bar\eta}$ being the dilatation~\cite{Emelyanov2}. These modes have been found 
in~\cite{Tanaka&Sasaki} for the upper Milne wedge which specify the Minkowski vacuum. One can also obtain the 
minkowskian modes in the lower wedge by analytically continuing them from the Rindler spacetime to that wedge. 
Thus, one has
\beqa\label{eq:milne-mink-modes}
\Phi_{p lm}(\bar{x}_{\pm}) &=& \pm i^le^{\pm\frac{\pi p}{2}}
\frac{e^{\mp ip\bar\eta_{\pm}}}{a(\bar\eta_{\pm})}
\frac{\Gamma\left(1 + l + ip\right)}{(4\pi\sinh\bar\chi_{\pm})^{\frac{1}{2}}}\;
P_{ip - \frac{1}{2}}^{-l-\frac{1}{2}}(\cosh\bar\chi_{\pm})Y_{lm}(\bar\Omega)\,, \;\; p \;\in\; \mathbf{R}\,,
\eeqa
where $\eta_{\pm}$ and $\chi_{\pm}$ are defined through $x^0 = \pm e^{\pm\bar\eta_{\pm}}\cosh\bar\chi_{\pm}$ and 
$|\mathbf{x}| = e^{\pm\bar\eta_{\pm}}\sinh\bar\chi_{\pm}$, where the plus and minus indices refer to the upper and 
lower Milne wedges, respectively.

By using Eq.~\eqref{eq:milne-mink-modes}, one can define conformal modes specifying the conformal Milne vacua 
in the wedges as follows
\bsubeqs\label{eq:mode-relation-milne}
\beqa
\Phi_{\omega lm}^{+}(x) &=& \alpha_{\omega lm}\Phi_{+\omega lm}(x)
+ \beta_{\omega lm}\Phi_{-\omega lm}^{*}(x)\,,
\\[1mm]
\Phi_{\omega lm}^{-}(x) &=& \alpha_{\omega lm}\Phi_{-\omega lm}^{*}(x)
+ \beta_{\omega lm}\Phi_{+\omega lm}^{}(x)\,,
\eeqa
\esubeqs
where $\omega \in \mathbf{R}^{+}$ and the Bogolyubov coefficients are
\beqa
\alpha_{\omega lm} &=& \frac{\exp\left(+\frac{\pi\omega}{2}\right)}{\left(2\sinh(\pi\omega)\right)^{\frac{1}{2}}}\,, \quad
\beta_{\omega lm} \;\;=\;\; -(-1)^l\frac{\exp\left(-\frac{\pi\omega}{2}\right)}{\left(2\sinh(\pi\omega)\right)^{\frac{1}{2}}}\,.
\eeqa
Note that the modes $\Phi_{\omega lm}^{+}(x)$ vanish in the lower wedge, while 
$\Phi_{\omega lm}^{-}(x)$ are strictly zero in the upper Milne wedge and both of them can be 
analytically continued into Rindler space.

One can express the Minkowski modes $\Phi_{p lm}(x)$ through the upper and lower conformal Milne ones, i.e. 
$\Phi_{\omega lm}^{\pm}(x)$, by exploiting Eq.~\eqref{eq:mode-relation-milne}. In other words, the Minkowski vacuum 
can be represented as an excited state under $|0^+\rangle\otimes|0^-\rangle$, where $|0^{\pm}\rangle$ is the conformal
Milne vacua in the upper and lower wedges, respectively. The crucial ingredient of the model 
that allows such representation is a presence of the conformal symmetry, because then the commutator 
$[\hat{\Phi}(x_1),\hat{\Phi}(x_2)]$ vanishes whenever $x_1$ and $x_2$ lie in the different Milne wedges as it has 
been noted in~\cite{Olson&Ralph}. Thus, the field degrees of freedom in one wedge are mutually independent from 
those lying in the other wedge.

\subsection{Local thermal observables}
\label{susec:lte-milne}

It is argued in~\cite{Buchholz&Ojima&Roos,Buchholz} that a real thermometer is modeled in the theoretical
language with the help of the Wick square of the field. This construction and a concept of the local equilibrium
have been further studied in curved spacetimes~\cite{Buchholz&Schlemmer,Solveen} (see also~\cite{SolveenDiss}).
Specifically, the squared value of the local temperature $\text{T}(x)$ measured by the thermometer in a given 
state $\omega$\footnote{It is a standard designation of a state in the algebraic approach to the quantum field
theory, see~\cite{Hollands&Wald&QFT,Wald}. I use it below for the convenience.} 
is probed by the Wick square as follows
\beqa
\text{T}^2(x) &=& 12\omega\big(\text{:}\hat{\Phi}^2(x)\text{:}\big)\,,  
\eeqa
wherein the Wick square is defined as
$\text{:}\hat{\Phi}(x_1)\hat{\Phi}(x_2)\text{:} \equiv \hat{\Phi}(x_1)\hat{\Phi}(x_2) - H(x_1,x_2)\hat{1}$, 
where $H(x_1,x_2)$ is the Hadamard parametrix canceling the divergent terms in the two-point function 
in the coincidence limit $x_2 \rightarrow x_1$~\cite{Wald}. 

The conformal KMS state $\omega_{\beta}$ specified by the KMS parameter $\beta$ (the inverse temperature) can 
be chosen as a thermal reference one. The two-point function in $\omega_{\beta}$ equals
\beqa
\omega_{\beta}(\hat{\Phi}(\bar{x}_1)\hat{\Phi}(\bar{x}_2)) &=& \frac{i}{2\pi a(\bar\eta_1)a(\bar\eta_2)}
\int\limits_{-\infty}^{+\infty}dk\;\frac{1}{1 - e^{-\beta k}}\int\limits_{-\infty}^{+\infty}d\bar\eta\;
\Delta_\text{r}(\bar\eta+\bar\eta_1;\bar\eta_2)\,e^{ik\bar\eta}\,,
\eeqa
where $\Delta_\text{r}(\bar{x}_1,\bar{x}_2)$ is the rescaled casual propagator, i.e. 
$[\hat{\Phi}_\text{r}(\bar{x}_1),\hat{\Phi}_\text{r}(\bar{x}_2)] = i\Delta_\text{r}(\bar{x}_1,\bar{x}_2)\hat{1}$ and
$\hat{\Phi}_\text{r}(\bar{x}) \equiv a(\bar\eta)\hat{\Phi}(\bar{x})$. Exploiting the previous Subsection, one finds
\beqa\label{eq:local-temperature-milne}
\text{T}^2(\bar{x}) &=& \frac{1}{a^2(\bar\eta)}\left(\frac{1}{\beta^2} - \frac{1}{4\pi^2}\right).
\eeqa
The Hadamard parametrix in this model coincides with the Minkowski two-point function, therefore, the local 
temperature in the Minkowski vacuum state is strictly zero. However, the Wick square in the conformal Milne vacuum 
is negative and coincides with the right-hand side of Eq.~\eqref{eq:local-temperature-milne} in the limit $\beta \rightarrow +\infty$, s.t. 
$\text{T}^2(\bar{x}) = -1/4\pi^2 a^2(\bar\eta)$.

The renormalized vacuum expectation value of the energy-momentum tensor $\hat{T}_{\nu}^{\mu}(x)$ vanishes 
in the Minkowski vacuum. Hence, one obtains
\beqa
\omega_{\beta}\big(\hat{T}_{\nu}^{\mu}(\bar{x})\big) &=& \frac{1}{480\pi^2a^4(\bar\eta)}
\left(\left(\frac{2\pi}{\beta}\right)^4-1\right)\left(\delta_{\nu}^{\mu}-\frac{4}{3}\delta_{i}^{\mu}\delta_{i\nu}\right),
\eeqa
wherein $i$ runs from 1 to 3. The thermal energy-momentum tensor $E_{\nu}^{\mu}(\bar{x})$ in this 
framework is, however, only a part of the total energy-momentum tensor $T_{\nu}^{\mu}(\bar{x})$ 
(denoted by $\epsilon_{\nu}^{\mu}$ in~\cite{Solveen}), such that it equals
\beqa
\omega_{\beta}\big(\hat{E}_{\nu}^{\mu}(\bar{x})\big) &=& \frac{1}{120\pi^2a^4(\bar\eta)}
\left(1+ \frac{4\pi^4}{\beta^4} - \frac{5\pi^2}{\beta^2}\right)
\left(\delta_{\nu}^{\mu}-\frac{4}{3}\delta_{i}^{\mu}\delta_{i\nu}\right).
\eeqa
Both $\omega_{\beta}\big(\hat{T}_{\nu}^{\mu}(\bar{x})\big)$ and $\omega_{\beta}\big(\hat{E}_{\nu}^{\mu}(\bar{x})\big)$
are traceless and vanish in the Minkowski vacuum, but $\omega_{\beta}\big(\hat{T}_{0}^{0}(\bar{x})\big)$ is less
then zero in the conformal Milne vacuum, while $\omega_{\beta}\big(\hat{E}_{0}^{0}(\bar{x})\big)$ is positive in it.
Note that neither $\omega_{\beta}\big(\hat{T}_{\nu}^{\mu}(\bar{x})\big)$ nor $\omega_{\beta}\big(\hat{E}_{\nu}^{\mu}(\bar{x})\big)$
are proportional to $\text{T}^4(\bar{x})$ as this the case for the pure thermal radiation. 

\section{Open anti-de Sitter spacetime}
\label{sec:ads}

Anti-de Sitter spacetime can be imagined as a four dimensional hyperboloid embedded in a five-dimensional
space $\mathbf{R}^5$ with the line element $ds^2 = \eta_{ab}dx^{a}dx^{b}$, where 
$a$ and $b$ run from 0 to 4 and $\eta_{ab} = \text{diag}(+,-,-,-,+)$, i.e. $\eta_{ab}x^ax^b = 1$.
This spacetime is pathological from the point of view that it is not globally hyperbolic. The first reason lies in
that its topology is $\mathbf{S}\times\mathbf{R}^3$, so that it possesses closed time-like curves. This feature
is cured by unwrapping the circle $\mathbf{S}$ and considering instead its universal covering $\mathbf{R}$.  
The second reason consists in that its spatial infinity is time-like. Therefore, one has to set a boundary 
condition there to fix the energy flux through it. These allow to have a well-defined quantum theory in the 
AdS hyperboloid~\cite{Avis&Isham&Storey}.

Among of possible parameterizations of the AdS hyperboloid, I shall consider so-called open
coordinates. The line element in these coordinates becomes
\beqa\label{eq:wedges}
ds^2 &=& a^2(\bar{\eta})\big(d\bar{\eta}^2 - d\bar{\chi}^2 - \sinh^2\bar{\chi} d\bar{\Omega}^2\big)\,,
\eeqa
where $a(\bar{\eta}) = 1/\cosh\bar{\eta}$. By the same metric one can describe geometry inside
the wedges for which $\eta \in (2\pi k, \pi + 2\pi k)$, where $\eta$ is the time coordinate in the static 
frame~\cite{Emelyanov} and $k \in \mathbf{Z}$. The geodesics of AdS space correspond to comoving
geodesics in open AdS space, i.e. integral curves of the conformal Killing vector $\xi = \partial_{\bar{\eta}}$, i.e.
the dilatation~\cite{Emelyanov2}.

For the analysis below it is needed to introduce coordinates that cover the rest part of the AdS hyperboloid.
They can be obtained by setting $\bar{\eta} = \tilde{\chi} - \frac{i\pi}{2}$ and $\bar{\chi} = \tilde{\eta} + \frac{i\pi}{2}$.
These coordinates correspond to a parametrization of the wedges with 
$\eta \in (-\frac{\pi}{2}+ 2\pi k, +\frac{\pi}{2} + 2\pi k)$, such that the line element equals
\beqa\label{eq:another-wedges}
ds^2 &=& a^2(\tilde{\chi})\big(d\tilde{\eta}^2 - d\tilde{\chi}^2 - \cosh^2\tilde{\eta} d\tilde{\Omega}^2\big)\,,
\eeqa
where $a(\tilde{\chi}) = 1/\sinh\tilde{\chi}$. Up to the scale factor it is similar to the Rindler 
universe in the spherical coordinates~\cite{Tanaka&Sasaki}. Therefore, the line element 
\eqref{eq:another-wedges} is conformally related with the Minkwoski one, specifically
\beqa\label{eq:metric}
ds^2 &=& \frac{4\,\eta_{\mu\nu}dx^{\mu}dx^{\nu}}{(1 + \eta_{\lambda\rho}x^{\lambda}x^{\rho})^2}\,,
\eeqa
where $x^{\mu} = (r\tanh\tilde{\eta},r\sin\theta\cos\varphi,r\sin\theta\sin\varphi,r\cos\theta)$ and 
$r = e^{\tilde{\chi}}\cosh\tilde{\eta}$, such that $t^2 - r^2 < 0$. If one takes $r^2 - t^2 > 0$, then this metric
can be transformed to the form it has in Eq.~\eqref{eq:wedges}. Thus, one can cover the AdS hyperboloid by domains being
conformally related to Minkowski space. Note that changing the sign inside the scale factor in~\eqref{eq:metric}, 
one obtains the de Sitter line element.

The Killing vector 
\beqa
\zeta &=& x^3\partial_{x^0} + x^0\partial_{x^3} \;=\;
\cos\tilde\theta\partial_{\tilde\eta} - \tanh\tilde\eta\sin\tilde\theta\partial_{\tilde\theta}
\eeqa
in those wedges is timelike for 
$\tilde\theta \in \{0,\pi\}$\footnote{It is not a restriction, because the rotation group SO(3) is a subgroup of 
the AdS symmetry group SO(2,3), so that one can always set those values of $\tilde\theta$ without loss of generality.}
and sets dynamics for an observer moving with a constant four-acceleration through AdS space, while 
$\sigma = \partial_{x^0}$ is the conformal Killing vector.\footnote{See~\cite{Emelyanov2} for more details.}
Due to the conformal symmetry, one can expand the rescaled field through the plane modes or through the boost 
ones, i.e. modes being eigenfunctions of $\sigma$ or the boost operator $\zeta$ in Minkowski space, respectively.
On the other hand, an observer moving along integral curves of $\zeta$ in AdS space defines Unruh modes up to the 
conformal factor. Hence, these vacua are thermally related as it has been first found in~\cite{Deser&Levin} by 
employing the Unruh-DeWitt detector~\cite{Birrell&Davies}.

\subsection{Static AdS modes in open AdS space}

As in the case of the Milne universe it is convenient to expand the field through the modes which when rescaled are 
eigenfunctions of $\xi$ and still define the static vacuum~\cite{Avis&Isham&Storey}. I argued in~\cite{Emelyanov} that these 
modes must exist. I show the modes below in this Subsection.

For the wedge $\eta \in (-\frac{\pi}{2},+\frac{\pi}{2})$ these modes are
\beqa\label{eq:static-modes}
\Phi_{plm}(\tilde{x}) &=& -i^{l+1}
\frac{e^{-ip\tilde{\chi}}}{a(\tilde{\chi})}\frac{\Gamma\left(1+l+ip\right)}{\left(4\pi i\cosh\tilde{\eta}\right)^{\frac{1}{2}}}
P_{ip-\frac{1}{2}}^{-l-\frac{1}{2}}\big(i\sinh\tilde{\eta}\big)Y_{lm}(\tilde{\Omega})\,, \;\; p \;\in\; \mathbf{R}
\eeqa
which are normalized on the effective Cauchy surface $\Sigma = \Sigma_1\cup\Sigma_2$,
i.e. these modes correspond to the ``transparent" boundary conditions~\cite{Avis&Isham&Storey}.

Performing an analytic continuation into the wedges $\eta \in (0,+\pi)$ and $\eta \in (-\pi,0)$
according to $\bar{\eta}_{+} = \tilde{\chi} - \frac{i\pi}{2}$, $\bar{\chi}_{+} = \tilde{\eta} + \frac{i\pi}{2}$ and
$\bar{\eta}_{-} = -\tilde{\chi} - \frac{i\pi}{2}$, $\bar{\chi}_{-} = \tilde{\eta} + \frac{i\pi}{2}$, respectively, one obtains
\beqa\label{eq:static-modes-pm}
\Phi_{plm}(\bar{x}_{\pm}) &=& 
\pm i^le^{\pm\frac{\pi p}{2}}\frac{e^{\mp p\bar{\eta}_{\pm}}}{a(\bar{\eta}_{\pm})}
\frac{\Gamma\left(1 + l + ip\right)}{(4\pi\sinh\bar{\chi}_{\pm})^{\frac{1}{2}}}\,
P_{ip-\frac{1}{2}}^{-l-\frac{1}{2}}(\cosh\bar{\chi}_{\pm})Y_{lm}(\bar\Omega_{\pm})\,.
\eeqa

Having these modes, it is straightforward to obtain the relation between the conformal and static vacua 
in AdS space~\cite{Emelyanov}. Indeed, the conformal modes are
\bsubeqs
\beqa
\Phi_{\omega lm}^{+}(x) &=& \alpha_{\omega lm}\Phi_{+\omega lm}(x) +
\beta_{\omega lm}\Phi_{-\omega lm}^{*}(x)\,,
\\[1mm]
\Phi_{\omega lm}^{-}(x) &=& \alpha_{\omega lm}\Phi_{-\omega lm}^{*}(x) +
\beta_{\omega lm}\Phi_{+\omega lm}(x)\,,
\eeqa
\esubeqs
where $\omega \in \mathbf{R}^{+}$ and 
\beqa
\alpha_{\omega lm} &=& \frac{\exp\left(+\frac{\pi\omega}{2}\right)}{(2\sinh(\pi\omega))^{\frac{1}{2}}}\,, \quad
\beta_{\omega lm} \;=\; -(-1)^l\frac{\exp\left(-\frac{\pi\omega}{2}\right)}{(2\sinh(\pi\omega))^{\frac{1}{2}}}\,.
\eeqa
One can show that the modes $\Phi_{\omega lm}^{+}(x)$ vanish in the wedges where $\eta \in (-\pi + 2\pi k, 2\pi k)$,
while the modes  $\Phi_{\omega lm}^{-}(x)$ vanish in the wedges where $\eta \in (2\pi k,+\pi + 2\pi k)$, 
$k \in \mathbf{Z}$.

The modes given in Eq.~\eqref{eq:static-modes} or Eq.~\eqref{eq:static-modes-pm} define the static vacuum. Indeed,
the two-point function is
\beqa\nonumber
\omega_\text{S}(\hat{\Phi}(x_1)\hat{\Phi}(x_2)) 
&=& \sum\limits_{lm}\int\limits_{-\infty}^{+\infty}dp\,
\Phi_{plm}^{\text{S}}(x_1)\Phi_{plm}^{\text{S}*}(x_2)\,
\;=\; \frac{1}{8\pi^2}
\frac{\cos\chi_1\cos\chi_2}{\cos(\Delta\eta - i\varepsilon) - \cos(\zeta)}\,,
\eeqa
where it has been already rewritten in the static AdS coordinates, 
$\cos\zeta = \cos(\chi_1 - \chi_2) + \sin\chi_1\sin\chi_2(\cos\Theta -1)$ and
$\cos\Theta = \cos\theta_1\cos\theta_2+\sin\theta_1\sin\theta_2\cos(\phi_1-\phi_2)$.
Comparing it with the two-point function for the closed Einstein static universe derived in~\cite{Birrell&Davies}
conformally mapped to AdS space, one concludes the modes given in Eq.~\eqref{eq:static-modes-pm} are 
unitary equivalent to the static ones for the ``transparent" boundary 
conditions~\cite{Avis&Isham&Storey}.\footnote{Note that the ``reflective" boundary conditions~\cite{Avis&Isham&Storey} are realized
by taking the modes $\mathbf{P_{\pm}}\Phi_{plm}(x)$, where $\mathbf{P}_{\pm} = \frac{1}{2}(\mathbf{1} \pm \mathfrak{I}_\text{C})$ and
$\mathfrak{I}_\text{C}$ is the so-called conformal inversion~\cite{Emelyanov2}.}

\subsection{Local thermal observables}
\label{susec:lte-oads}

In analogous manner to Subsec. \ref{susec:lte-milne}, one finds the squared value of the local temperature ascribed to
the conformal KMS state with the inverse temperature $\beta$:
\beqa\label{eq:local-temperature-ads}
\text{T}^2(\bar{x}) &=& \frac{1}{a^2(\bar{\eta})}\left(\frac{1}{\beta^2} - \frac{1}{4\pi^2}\right) + \frac{R}{24\pi^2} - 12\alpha_0R\,,
\eeqa
where $\alpha_0$ is due to ambiguity in defining the Wick square~\cite{Hollands&Wald} and 
$R = -6(a^{\prime\prime}/a - 1)/a^2$ Ricci scalar equaling to $+12$ in AdS space. If one sets
$\alpha_0 = 1/288\pi^2$, then the Wick square is a conformally invariant field~\cite{Pinamonti}(see also~\cite{SolveenDiss}).
Another motivation for this choice of $\alpha_0$ is given in~\cite{Solveen,SolveenDiss}. This value of $\alpha_0$ is taken
for granted in this Subsection.

The AdS vacuum restricted to open AdS space is a conformal KMS state with $\beta = 2\pi$. The squared local temperature 
of the conformal vacuum is negative and equals $-1/4\pi^2a^2(\bar{\eta})$.

Taking into account the renormalized vacuum expectation value of $\hat{T}_{\nu}^{\mu}(\bar{x})$ in the AdS vacuum in open AdS
space, one derives
\beqa\label{eq:emt-ads-sv}
\omega_{\beta}\big(\hat{T}_{\nu}^{\mu}(\bar{x})\big) &=& \frac{1}{960\pi^2}\,\delta_{\nu}^{\mu} + \frac{1}{480\pi^2a^4(\bar{\eta})}
\left(\left(\frac{2\pi}{\beta}\right)^4-1\right)\left(\delta_{\nu}^{\mu}-\frac{4}{3}\delta_{i}^{\mu}\delta_{i\nu}\right),
\eeqa
whereas for the thermal energy-momentum tensor $\hat{E}_{\nu}^{\mu}(\bar{x})$ one finds an expression to be
structurally rather different from the right-hand side of Eq.~\eqref{eq:emt-ads-sv}.

\section{Discussion}
\label{sec:discussion}

An analogous result to~\eqref{eq:local-temperature-ads} one obtains for the local temperature squared in open de 
Sitter spacetime, where $R = -12$ and $a(\bar{\eta}) = 1/\sinh\bar{\eta}$ in that equation. The total renormalized
energy-momentum tensor $\hat{T}_{\nu}^{\mu}(\bar{x})$ in the conformal KMS state $\omega_{\beta}$ is functionally given by 
the same Eq.~\eqref{eq:emt-ads-sv}. The conformal KMS state with $\beta = 2\pi$ corresponds to the conformal or 
Chernikov-Tagirov vacuum~\cite{Birrell&Davies} defined on the whole dS hyperboloid.

As has been mentioned above, the Wick square is ambiguous~\cite{Hollands&Wald}. This means one has to impose an
extra condition to get rid of that. The value of $\alpha_0 = 1/288\pi^2$ has been motivated in~\cite{Solveen,SolveenDiss}.
However, one has to set a zero value of $\alpha_0$ to have a zero local temperature for an observer freely moving along
the time translation Killing vector in the open Einstein static universe. Indeed, the local temperature squared can be immediately
obtained from Eq.~\eqref{eq:local-temperature-ads} by setting $a(\bar{\eta}) = 1$, so that if the quantum field is in the static ESU 
vacuum, then $\beta \rightarrow +\infty$. Hence, $\text{T}^2(\bar{x}) = 0$ if and only if $\alpha_0 = 0$, where $R = +6$ in the open 
Einstein static universe has been taken into account. The renormalized expectation values of the total and thermal 
energy-momentum tensors in the KMS state defined with respect to the Killing vector $\partial_{\bar\eta}$ coincide and are 
equal to
\beqa
\omega_{\beta}\big(\hat{T}_{\nu}^{\mu}(\bar{x})\big) &=&
\omega_{\beta}\big(\hat{E}_{\nu}^{\mu}(\bar{x})\big) \;=\; 
\frac{\pi^2}{60\beta^4}\left(\delta_{\nu}^{\mu}-\frac{4}{3}\delta_{i}^{\mu}\delta_{i\nu}\right).
\eeqa
Note that they are both proportional to $\text{T}^4(\bar{x})$ for vanishing $\alpha_0$. This result then is similar to that in 
Minkowski space for an inertial observer, wherein, however, the value of $\alpha_0$ is irrelevant, because the scalar 
curvature vanishes.

As has been noted above, the Chernikov-Tagirov state in closed de Sitter space is also the conformal vacuum
defined with respect to the conformal Killing vector $\partial_{\eta}$. The local temperature squared is 
\beqa\label{eq:diss-lt-closed}
\text{T}^2(x) &=& \frac{6}{\pi^2a^2(\eta)}\sum\limits_{n = 0}^{+\infty}\frac{n}{e^{\beta n} -1} + \frac{R}{24\pi^2} - 12\alpha_0R\,,
\eeqa
where $a(\eta) = 1/\sin\eta$ and $R = -6(a^{\prime\prime}/a + 1)/a^2$ Ricci scalar equaling to $-12$
in dS space. This result is a slight generalisation of that obtained in~\cite{Stottmeister}. 
The expectation value of $\hat{T}_{\nu}^{\mu}(x)$ in the conformal KMS state defined with respect to $\partial_{\eta}$ is given by 
\beqa\label{eq:cds-emt}
\omega_{\beta}\big(\hat{T}_{\nu}^{\mu}(x)\big) &=& \frac{1}{960\pi^2}\,\delta_{\nu}^{\mu} + \frac{1}{2\pi^2a^4(\eta)}
\sum\limits_{n = 0}^{+\infty}\frac{n^3}{e^{\beta n} -1}\left(\delta_{\nu}^{\mu}-\frac{4}{3}\delta_{i}^{\mu}\delta_{i\nu}\right),
\eeqa
where the renormalized vacuum expectation value of $\hat{T}_{\nu}^{\mu}(x)$ in the Chernikov-Tagirov state has been taken 
into account~\cite{Birrell&Davies}, while $\omega_{\beta}\big(\hat{E}_{\nu}^{\mu}(x)\big)$ has structurally a different form in 
comparison with $\omega_{\beta}\big(\hat{T}_{\nu}^{\mu}(x)\big)$ given in Eq.~\eqref{eq:cds-emt}. 

A comoving observer in closed dS space moving along curves with the tangent vector $\partial_{\eta}$
has to register in the standard interpretation a thermal bath with the Gibbons-Hawking 
temperature $\text{T}_\text{GH} = 1/2\pi$~\cite{Gibbons&Hawking}. This implies $\alpha_0 = 1/192\pi^2$. 
However, the local temperature is zero if $\alpha_0 = 1/288\pi^2$~\cite{Solveen,SolveenDiss}.

These energy-momentum tensors, i.e. $\hat{T}_{\nu}^{\mu}(x)$ and $\hat{E}_{\nu}^{\mu}(x)$, in the thermal state coincide in the 
closed Einstein static universe. The local temperature squared can be obtained from Eq.~\eqref{eq:diss-lt-closed} by setting 
$a(\eta) = 1$, such that $R = -6$ and $\alpha_0 = 1/288\pi^2$~\cite{SolveenDiss}.  Employing the renormalized vacuum 
expectation value of $\hat{T}_{\nu}^{\mu}(x)$ for the closed ESU~\cite{Birrell&Davies,Candelas&Dowker}, one obtains
\beqa
\omega_{\beta}\big(\hat{T}_{\nu}^{\mu}(x)\big) &=&
\omega_{\beta}\big(\hat{E}_{\nu}^{\mu}(x)\big) \;=\; 
\frac{1}{2\pi^2}\left(
\sum\limits_{n = 0}^{+\infty}\frac{n^3}{e^{\beta n} -1} 
+ \frac{1}{240}\right)\left(\delta_{\nu}^{\mu}-\frac{4}{3}\delta_{i}^{\mu}\delta_{i\nu}\right),
\eeqa
which are clearly not proportional to the quartic value of the local temperature as in the case of the pure thermal radiation.

\section{Concluding remarks}
\label{sec:conclusions}

In the present paper I have considered certain local thermal observables in the well-known examples of spatially open 
FRW spaces for the conformal linear field theory. These observables have been put forward in a series of 
articles~\cite{Buchholz&Ojima&Roos,Buchholz,Buchholz&Schlemmer,Solveen,SolveenDiss}
with a goal to define a local thermal equilibrium under the influence of external fields, in particular, in
curved spacetimes.

I have found that the ambiguity in defining the Wick square parametrized by $\alpha_0$~\cite{Hollands&Wald}
has no universal value, i.e. depends on a particular situation. It has to be zero in the open Einstein static universe, 
otherwise the physical meaning of the local temperature $\text{T}(x)$ as defined in~\cite{Buchholz&Ojima&Roos,Buchholz} 
is lost.

Assuming that $\alpha_0$ has to be zero for all open FRW universes considered above, one is forced to conclude
that the local temperature of the Minkowski vacuum vanishes, but is real for the AdS vacuum restricted to open AdS space 
and imaginary for the Chernikov-Tagirov state restricted to open dS space. Thus, the physical meaning of the local 
temperature as a model for a real thermometer becomes lost in the last case.

The local temperature for the conformal vacua in Milne and open dS spaces are imaginary, but is real only for a short 
interval of $\bar{\eta}$ close to $\bar{\eta} = 0$ in open anti-de Sitter space. However, the backreaction of the field 
being in the conformal vacuum is infinite on horizons, so that these vacua cannot anyway be physically realized. 

I have considered in~\cite{Emelyanov} spacetime that approaches the open Einstein static universe at future- and 
past-time infinities and has a phase when space looks like open anti-de Sitter spacetime. During the AdS phase one might 
await that a comoving observer detects a thermal radiation with temperature $\text{T} = (2\pi a(\bar{\eta}))^{-1}$ 
which vanishes at $\bar{\eta} \rightarrow \pm\infty$~\cite{Emelyanov}. Qualitatively it behaves itself as the local 
temperature $\text{T}(x)$ around $\bar\eta = 0$ and in the limits $\bar{\eta} \rightarrow \pm\infty$. However, 
$\text{T}^2(x)$ still has no physical sense, because it possesses negative values. 

To sum it up, the concept of the local temperature as originally defined is at least questionable. Perhaps, its
appropriately modified version could be identified with the readings of a real thermometer. This demands, however, 
further investigations.

\section*{%\hspace*{-4.5mm}
ACKNOWLEDGMENTS}

This research is supported by TRR 33 ``The Dark Universe''.

\end{document}